\documentclass{jetpl-160406060-v3}

\twocolumn  

\lat

\title{Brane realization of $q$-theory and the cosmological constant problem}

\rtitle{Brane realization of $q$-theory}

\sodtitle{Brane realization of $q$-theory and cosmological constant problem}

\author{F.R. Klinkhamer$^{+}$\/\thanks{e-mail: frans.klinkhamer@kit.edu},
G.E. Volovik$^{*\&}$\/\thanks{e-mail: volovik@boojum.hut.fi}}

\rauthor{F.\,R.\,Klinkhamer, G.\,E.\,Volovik}

\sodauthor{Klinkhamer, Volovik}

\address{
$^+$Institute for
Theoretical Physics, Karlsruhe Institute of
Technology (KIT), 76128 Karlsruhe, Germany\\~\\
$^*$Low Temperature Laboratory, Aalto University,
P.O. Box 15100, FI-00076 Aalto, Finland\\~\\
$^\&$ Landau Institute for Theoretical Physics RAS, Kosygina 2,
119334 Moscow, Russia
\vspace*{5mm}\newline
}

\abstract{
We discuss the cosmological constant problem using the properties of a freely-suspended two-dimensional condensed-matter film, i.e.,
an explicit realization of a 2D brane.
The large contributions of vacuum fluctuations to the surface tension
of this film are cancelled in equilibrium by the thermodynamic potential
arising from the conservation law for particle number.
In short, the surface tension of the film
vanishes in equilibrium due to a thermodynamic identity.
This 2D brane can be generalized to a 4D brane with gravity.
For the 4D brane, the analogue of the 2D surface tension is the
4D cosmological constant, which is also nullified in full equilibrium.
The 4D brane theory provides an alternative description of the phenomenological $q$-theory of the quantum vacuum.
As for other realizations of the vacuum variable $q$, such as the 4-form field-strength realization, the main ingredient is the conservation law for the variable $q$, which makes the vacuum
a self-sustained system.
For a vacuum within this class, the nullification of the cosmological constant takes place automatically in equilibrium.
Out of equilibrium, the cosmological constant can be as large as suggested by naive estimates based on the summation of zero-point energies. In this brane description, $q$-theory also corresponds to a generalization of unimodular gravity.
}


\begin{document}
\maketitle

\section{Introduction}
\label{sec:Intro}

General relativity and the relativistic quantum field theory
of the Standard Model
are probably effective theories, as manifested by their ultraviolet divergences.
Especially troublesome is the quartic divergence of the energy density
from the vacuum fluctuations,
which leads to the so-called cosmological constant
problem (CCP)
~\cite{Weinberg1989,Weinberg1996,Nobbenhuis2006}.

A particular mechanism for the nullification of
the relevant vacuum energy density works for vacua
which have the property of being
\emph{self-sustained media}~\cite{KlinkhamerVolovik2008a,%
KlinkhamerVolovik2008b,%
KlinkhamerVolovik2009a,KlinkhamerVolovik2009b,%
KlinkhamerVolovik2010}.
A self-sustained medium has a definite macroscopic
volume even in the absence of an environment. Two condensed-matter
examples are a droplet of quantum liquid at zero temperature in empty space~\cite{Volovik2003} and a freely-suspended film~\cite{Kats2015}.
Here, we focus on the last example.

The Lorentz-invariant self-sustained medium relevant to the CCP
is characterized by a relativistic scalar $q$. Distinct from a fundamental scalar
$\phi$, the vacuum variable $q$ allows the medium to exist without external environment.

\section{Fundamental vs. conserved scalars}
\label{sec:Fundamental-vs-conserved}

In order to understand the difference between fundamental and conserved scalars, let us compare gravity with a fundamental scalar $\phi$ and gravity with a conserved scalar $q$ obtained from a 4-form field strength $F$
\cite{DuffNieuwenhuizen1980,Aurilia-etal1980,Hawking1984,Duff1989,
DuncanJensen1989,BoussoPolchinski2000,Aurilia-etal2004,Wu2008},
with
$F_{\kappa\lambda\mu\nu} \propto q\, \epsilon_{\kappa\lambda\mu\nu}$
and
$q^2 \propto F_{\kappa\lambda\mu\nu}\,F^{\kappa\lambda\mu\nu}$.
Let us consider these theories for the simplest possible arrangement: no explicit derivatives of $\phi$ and $q$, and no direct couplings
of $\phi$ and $q$ to the Ricci scalar $R$.

The action for gravity with a non-dynamic fundamental scalar $\phi$
is
\begin{equation}
I_{1}=
-\int_{\mathbb{R}^4}\,d^4x\; \sqrt{-g}\,
\left(\frac{R}{16\pi G_N} +\epsilon_1(\phi)\right) \,.
\label{eq:action-phi}
\end{equation}
The action for gravity with a three-form gauge field $A$ is
\begin{equation}
I_{2}=
-\int_{\mathbb{R}^4}\,d^4x\; \sqrt{-g}\,
\left(\frac{R}{16\pi G_N} +\epsilon_2(q) \right) \,.
\label{eq:action-q}
\end{equation}
Both functions $\epsilon_1(\phi)$ and $\epsilon_2(q)$
may include a genuine cosmological constant $\Lambda_\text{bare}\,$.

Variation  over $g^{\mu\nu}$ of the action $I_{j}$, for $j=1$ or $2$, gives
the Einstein equation in standard form,
\begin{equation}
\frac{1}{8\pi G_{N}}
\left( R_{\mu\nu}-\frac{1}{2}\,R\,g_{\mu\nu}\right)=
\Lambda_{j}\; g_{\mu\nu}\,,
\label{eq:Einstein-eq-phi-or-q}
\end{equation}
but with different expressions for the vacuum energy density,
\begin{subequations} 
\label{eq:Lambda-phi-q}
\begin{eqnarray}
 \label{eq:Lambda-phi}
\Lambda_1(\phi) &=& \epsilon_1(\phi)\,,
\\[2mm]
 \label{eq:Lambda-q}
\Lambda_2(q)    &=& \epsilon_2(q) - q\,\frac{d\epsilon_2(q)}{dq} \,.
\end{eqnarray}
\end{subequations}
The calculation of the energy-momentum tensor
gives an extra term in the $\Lambda(q)$ expression above, because
the metric also enters the composite field $q$.
Recall, in fact, the precise definitions $q^2 \equiv
- (1/24)\,F_{\kappa\lambda\mu\nu}\,F^{\kappa\lambda\mu\nu}$
and
$F_{\kappa\lambda\mu\nu} \equiv \nabla_{[\kappa}A_{\lambda\mu\nu]}$,
with the three-form gauge field $A$,
the covariant derivative $\nabla_{\kappa}$,
and a pair of square brackets
around spacetime indices standing for complete anti-symmetrization.

The field equations for $\phi$ and $q$ are also different,
\begin{subequations} \label{eq:fieldEq-phi-q}
\begin{eqnarray}
\label{eq:fieldEq-phi}
\frac{d\epsilon_1(\phi)}{d\phi} &=&0 \,,
\\[2mm]
\label{eq:fieldEq-q}
\nabla_\nu \left(\frac{d\epsilon_2(q)}{dq} \right) &=& 0 \,.
\end{eqnarray}
\end{subequations}
The field equation for $\phi$ does not have the freedom to nullify the cosmological constant. The two
requirements for nullification,  $\epsilon_1(\phi)=0$ and $d\epsilon_1(\phi)/d\phi=0$, demand fine-tuning of the parameters entering the function $\epsilon_1(\phi)$,
cf. Refs.~\cite{Weinberg1989,Weinberg1996}.

In contrast, the solution of the field equation for $q$
is given by
\begin{equation}
\frac{d\epsilon_2(q)}{dq}=\mu
 \,,
\label{eq:fieldEqq}
\end{equation}
where $\mu$ is an arbitrary integration constant.
Hence, we have the complete freedom  of choosing the value
of $\mu$ in \eqref{eq:fieldEqq} as
regards solving the field Eq.~\eqref{eq:fieldEq-q},
which leaves us to consider the vacuum energy.
Writing $\Lambda_2(q)$ from Eq.~(\ref{eq:Lambda-q}) as
$\epsilon_2(q) - \mu\, q$, we then see that the huge vacuum energy density
stored in $\epsilon_2(q)$ can
be compensated by the counterterm $- \mu\, q$. This compensation occurs
due to thermodynamic identities applied to the equilibrium self-sustained vacuum (we shall see this for the example of a freely-suspended film in Sec.\,3\,).    
Here, $\mu$ plays the role of a chemical                   
potential~\cite{KlinkhamerVolovik2008b}, which is self-tuned 
in the equilibrium state of the quantum vacuum. The self-tuning gives rise to the nullification
of the vacuum energy density for the equilibrium vacuum in the absence of matter fields,
\begin{equation}
\Lambda_2(q_{\rm eq}) = \epsilon_2(q_{\rm eq}) - \mu_{\rm eq}\, q_{\rm eq}
= 0 \,.
\label{eq:Lambda-2-nullification}
\end{equation}
No such compensation is expected for the vacuum of the
fundamental scalar field $\phi$. However, if both $\phi$ and $q$ fields are present, then the energy $\epsilon_1(\phi)$
will also be compensated in the equilibrium vacuum by the readjustment of the chemical potential $\mu$~\cite{KlinkhamerVolovik2008a}.

\section{Freely suspended film}
\label{sec:Film}

The thin film considered here is a two-dimensional (2D) object embedded into the three-dimensional (3D) Euclidean space $\mathbb{R}^3$
with Cartesian coordinates $X,\, Y,\,Z$.
In modern parlance, this film is a 2D brane~\cite{Rubakov2001}.
We are interested in a freely-suspended film, which is
not surrounded by dense matter. In equilibrium, the film is parallel to
the $(X,Y)$-plane,
and this state is analogous to the 4D equilibrium Minkowski vacuum
to be discussed in Sec.~\ref{sec:4D-Brane}.

The Hamiltonian~\cite{Kats2015}  which describes the deformations of the film -- variation of the density of the film
and bending (displacement) of the film -- can be written in the following form:
\begin{equation}
H= \int d^2x\, \left[\sqrt{g}\;\epsilon\left(\frac{n}{\sqrt{g}}\right)
- \mu \,  n\right]+H_{\rm bending}  \,.
\label{FilmHamiltonian}
\end{equation}
Here, $n(x,y)\equiv \int dz \,n(x,y,z)$
is the 2D particle density obtained by integrating
over the  extra dimension of the bulk space $\mathbb{R}^3$.
For simplicity, we assume the absence of folding.
The total number of particles in the film, $N=\int d^2x \,n(x,y)$,
is a conserved quantity. We introduce
the corresponding Lagrange multiplier $\mu$,
which plays the role of a chemical potential. Particle conservation (or mass conservation in Ref.~\cite{Kats2015}) is
the main condition for the existence of a stable freely-suspended film.
Hence, the film belongs to the class of self-sustained systems.

The potential term [the first term in the integrand of
Eq.~(\ref{FilmHamiltonian})]
depends on the quantity $n/\sqrt{g}$,
the particle density per unit area of the curved film
(recall $dS=\sqrt{g}\,dxdy$).
Here and in Eq.~\eqref{FilmHamiltonian},
we have used the definition $g \equiv \det(g_{ik})$
for the curved-film metric $g_{ik}$ with signature ($+,\,+$).
The second term on the
right-hand side of Eq.~(\ref{FilmHamiltonian}) is the bending energy.
It contains gradients of the metric, but, due to the lack of invariance under general coordinate transformations,
it is not equivalent to the standard curvature term in gravitation theories. Still, it includes the Gauss curvature $R$, which is a total derivative in two dimensions.
Here, we are not really interested in the second term
$H_{\rm bending}$ of Eq.~(\ref{FilmHamiltonian}), since the central point of our argument will concern the first term, which will be seen to be equivalent to the vacuum energy. So, let us consider only the first term on the
right-hand side of Eq.~(\ref{FilmHamiltonian}).

Variation of the Hamiltonian \eqref{FilmHamiltonian}
over $n$ gives the following equation:
\begin{subequations} 
\label{eq:variation-2D-brane-mu-q}
\begin{eqnarray}
\label{eq:variation-2D-brane-mu}
\frac{d\epsilon}{dq}&=&\mu \,,
\\[2mm]
\label{eq:variation-2D-brane-q}
q                   &\equiv&\frac{n}{\sqrt{g}}\,,
\end{eqnarray}
\end{subequations}
and variation over $\sqrt{g}$ gives the surface tension
$\sigma$ of the film,
\begin{equation}
\sigma\equiv \frac{\delta H}{\delta \sqrt{g}}= \epsilon(q) -q\,\frac{d\epsilon}{dq}=\epsilon(q) -\mu\, q\,.
\label{eq:variationg}
\end{equation}
For the freely-suspended film (i.e., no forces from the environment),
the surface tension is zero in equilibrium,
$\sigma_{\rm eq}=0$. The chemical potential $\mu$ is self-tuned to reach the equilibrium state.
Since the variation over the metric $g^{ik}$ gives the stress tensor $T_{ik}=-\sigma g_{ik}$~\cite{Kats2015,Lebedev1989,Kats1993}, the surface tension $\sigma$ plays the role of a cosmological constant in the 2+0 gravity theory of the film, which is nullified in equilibrium.

As emphasized in Ref.~\cite{Kats2015}, the equilibrium condition $\sigma=0$ is not disturbed by quantum fluctuations,
because it is the consequence of the thermodynamic identity $\sqrt{g}\,\epsilon - \mu\, n= -P$, where $P$ is the external pressure
(thermal effects are not considered). In the absence of external forces, the surface tension is zero irrespective of the quantum fluctuations. Of course, the quantum fluctuations contribute to the energy $\epsilon$, and this contribution can be essential. But this contribution is always fully compensated by the counterterm $-\mu\, n$ in equilibrium,
which is, in fact, the property of any self-sustained vacuum.
This suggests that, if the vacuum of our Universe belongs to the class of self-sustained systems, its energy in equilibrium is fully cancelled in spite of the huge effects of vacuum fluctuations.
The cancellation results in a zero value for the cosmological constant in an equilibrium Universe without ponderable matter.

\section{4D brane}
\label{sec:4D-Brane}

The corresponding modification of
the Einstein action on the four-dimensional (4D) ``brane'' is
\begin{eqnarray}
\hspace*{-10mm}
I &=&- \int
d^4x\, \sqrt{-g}\,\left[\epsilon\left(\frac{n}{\sqrt{-g}}\right)+\frac{R}{16\pi G_{N}} +\mathcal{L}^{M}[\psi]\right]
\nonumber\\[2mm]
&&+ \mu \int\,d^4x \;n \,,
\label{EinsteinAction4D}
\end{eqnarray}
where the metric $g_{\mu\nu}$ has a Lorentzian signature
($-,\,+,\,+,\,+$) making
for a negative determinant $g \equiv \det(g_{\mu\nu})$
and where
$n$ is the 4D analog of the particle density of the 2D film
(the 4D density $n$ perhaps refers to the ``atoms''
of spacetime).
In principle, the gravitational coupling $G$ may also depend on $n$,
$G=G(n)$,but we fix $G=G_N$,  for simplicity.
Similarly, we omit any $n$-dependence
of the parameters in the matter Lagrange density
$\mathcal{L}^{M}(\psi)$, where $\psi$ stands for a generic matter
field.

Variation of the action \eqref{EinsteinAction4D}
over the density $n$ gives the analog of Eq.~(\ref{eq:variation-2D-brane-mu-q}),
\begin{subequations}
\label{eq:variation-4D-brane-mu-q}
\begin{eqnarray}
\label{eq:variation-4D-brane-mu}
\frac{d\epsilon}{dq}&=& \mu \,,
\\[2mm]
\label{eq:variation-4D-brane-q}
q                   &\equiv& \frac{n}{\sqrt{-g}}\,.
\end{eqnarray}
\end{subequations}
Variation of the action \eqref{EinsteinAction4D}
over the inverse metric $g^{\mu\nu}$ gives the Einstein equation,
\begin{equation}
\frac{1}{8\pi G_{N}}
\left( R_{\mu\nu}-\frac{1}{2}\,R\,g_{\mu\nu}\right)=
\Lambda(q)\, g_{\mu\nu}+T^{M}_{\mu\nu}\,,
\label{eq:Einstein-eq-4D-brane}
\end{equation}
with the following vacuum energy density:
\begin{equation}
\Lambda(q)= \epsilon(q) - q\,\frac{d\epsilon(q)}{dq}
=  \epsilon(q)  - \mu\, q\,,
\label{eq:Lambda}
\end{equation}
where the last equality relies on result \eqref{eq:variation-4D-brane-mu}.

The same equations can be obtained if the action (\ref{EinsteinAction4D}) is expressed  directly in terms of $q$, and if one varies  over $q$ and $g^{\mu\nu}$,  assuming that $q$ is a conserved quantity. This demonstrates the universal and generic properties of $q$-theory~\cite{KlinkhamerVolovik2008a},
which do not depend
on the particular realization of the conserved quantity $q$.

\section{Discussion}
\label{sec:Discussion}

In this Letter, we have introduced an alternative description of $q$-theory, which is based on an analogy with the two-dimensional (2D) condensed-matter film. The freely-suspended film belongs to the
class of self-sustained systems, which is characterized by a conservation law. For the case of a nonrelativistic 2D film,
the relevant conservation law is the conservation of mass of the film. Due to this conservation law, the surface tension of the film
is nullified in equilibrium, in spite of the large contributions of the vacuum fluctuations to the energy of the film.
For the (3+1)-dimensional relativistic analog of such a film,
the same type of conservation law leads to the nullification of
the cosmological constant -- the 3+1 analog of the 2D surface tension.

The 4D brane provides a new realization of $q$-theory. 
Recall that the main input of $q$-theory is the assumption that the quantum vacuum belongs to the class of
self-sustained systems, characterized by the conservation law for some kind of ``particle number.''
For a vacuum within this class, the nullification of the cosmological constant automatically takes place in equilibrium,
while, out of equilibrium, the cosmological constant can be large and comparable with the estimates
based on the summation of the zero-point energies of the quantum fields
(with appropriate cutoffs).

Since the quantity $\sqrt{-g}$ enters the action as a separate variable, this 4D-brane realization of $q$-theory represents a generalization of  unimodular gravity~\cite{Bij1982,Zee1983,Dragon1988,Henneaux1989,Weinberg1989}.
Gravity in four dimensions is, of course, very different from
gravity in two dimensions, for example, there are gravitational
waves (gravitons) in four dimensions but not in two.
For the 4D theory \eqref{EinsteinAction4D}, we may then consider
gravitational processes at energies far below
the binding energy of the  ``atoms'' of spacetime responsible for
the number density $n$ . For such low-energy processes,
$n$ is effectively fixed and nondynamical, $n=n_0$.
From the 4D general covariance of \eqref{EinsteinAction4D}, we have that
$n$ is a scalar density with the same weight as the square root of the
negative of the determinant of the metric. Introducing a prior metric
with determinant $g_0$,
we then have $n=n_0\propto \sqrt{-g_0}$     
and the $q$ variable is effectively        
equal to $q\propto \sqrt{-g_0}/\sqrt{-g}$. 
The theory \eqref{EinsteinAction4D} written in terms of
the inverse variable $\widetilde{\sigma} = \sqrt{-g}/\sqrt{-g_0}$  
is essentially the one studied in Ref.~\cite{Klinkhamer2016},  
where the role of vacuum-matter energy exchange has been investigated.

Now, return to the original form of $q$ as given by
Eq.~\eqref{eq:variation-4D-brane-q}.
It can then be shown that, for homogeneous matter fields in a cosmological context, the vanishing covariant divergence of
the vacuum-energy term on the right-hand side of
Eq.~\eqref{eq:Einstein-eq-4D-brane}
gives $\partial_t\,\Lambda = 0$, which implies $dq/dt=0$.
From the definition of $q$, we can then relate the rate of change of the
metric determinant to the rate of change of the brane number density,
$d(\ln \sqrt{-g})/dt=d(\ln n)/dt$.
The role of this type of intra-brane dynamics needs to be clarified.

Let us also comment on the main difference between our approach and
the one of Ref.~\cite{Sorkin2013}, where the cosmological constant was estimated as $\sqrt{N}$, with $N$ the number of elements.
Reference~\cite{Sorkin2013} noted that
the individual contributions to the action $I$
have random signs and assumed that their sum vanishes on average,
with residual fluctuations of order $\sqrt{N}$ being responsible for the observed value of $\Lambda$. In our case, the contribution to the action $I$ is proportional to $N$, while the cancellation
takes place for the quantity $\Lambda$ which enters the Einstein equations. $\Lambda$ does not necessarily coincide
with the vacuum energy  density $\epsilon$, which enters the action. The difference between $\Lambda$ and $\epsilon$
reflects two different definitions of the energy of quantum
fields on an external time-independent background, see Ref.~\cite{Fursaev1999}. The first
one defines the energy in terms of the stress-energy tensor, while the second one  identifies the energy with the Hamiltonian.

According to the $q$-theory approach to the cosmological constant
problem, the present small value of $\Lambda$ is the result
of the incomplete cancellation of the vacuum energy in a slowly evolving nonequilibrium Universe~\cite{KlinkhamerVolovik2008b}.

GEV thanks Efim Kats and Vladimir Lebedev for valuable discussions.

\textbf{Note Added in Proof.}   
An early paper on the vanishing surface tension of fluid membranes
as an analog of the near-zero cosmological constant of general relativity
is Ref.~\cite{SamuelSinha2006}.



\begin{thebibliography}{99}


\bibitem{Weinberg1989}
S. Weinberg,
``The cosmological constant problem,''
Rev. Mod. Phys.  {\bf 61}, 1 (1989).

\bibitem{Weinberg1996}
S. Weinberg,
``Theories of the cosmological constant,''
in: N. Turok, \emph{Critical Dialogues in Cosmology}
(World Scientific, Singapore, 1997), p. 195,
arXiv:astro-ph/9610044.

\bibitem{Nobbenhuis2006}
S. Nobbenhuis,
``Categorizing different approaches to the cosmological constant problem,''
Found.  Phys.  \textbf{36}, 613 (2006),
arXiv:gr-qc/0411093.


\bibitem{KlinkhamerVolovik2008a}
F.R. Klinkhamer and G.E. Volovik,
``Self-tuning vacuum variable and cosmological constant,''
Phys. Rev. D \textbf{77}, 085015 (2008), arXiv:0711.3170.


\bibitem{KlinkhamerVolovik2008b}
F.R. Klinkhamer and G.E. Volovik,
Dynamic vacuum variable and equilibrium approach in cosmology,
Phys. Rev. D \textbf{78}, 063528 (2008), arXiv:0806.2805.


\bibitem{KlinkhamerVolovik2009a}  
F.R. Klinkhamer and G.E. Volovik,
``Gluonic vacuum, $q$-theory, and the cosmological constant,''
Phys. Rev.  D {\bf 79}, 063527 (2009),
arXiv:0811.4347.

\bibitem{KlinkhamerVolovik2009b}  
F.R. Klinkhamer and G.E. Volovik,
``Vacuum energy density kicked by the electroweak crossover,''
Phys. Rev. D {\bf 80}, 083001 (2009),
arXiv:0905.1919.

\bibitem{KlinkhamerVolovik2010}
F.R.~Klinkhamer and G.E.~Volovik,
``Towards a solution of the cosmological constant problem,''
JETP Lett.\  {\bf 91}, 259 (2010),
arXiv:0907.4887.  


\bibitem{Volovik2003}
G.E. Volovik,
\emph{The Universe in a Helium Droplet}, Paperback Edition
(Oxford UP, 2008).

\bibitem{Kats2015}
E.I. Kats and V.V. Lebedev,
``Nonlinear fluctuation effects in dynamics of freely suspended films,''
Phys. Rev. E  \textbf{91}, 032415 (2015),
arXiv:1501.06703.  

\bibitem{DuffNieuwenhuizen1980}
M.J. Duff and P. van Nieuwenhuizen,
``Quantum inequivalence of different field representations,''
Phys. Lett.  B {\bf 94}, 179 (1980).

\bibitem{Aurilia-etal1980}
A. Aurilia, H. Nicolai, and P.K. Townsend,
``Hidden constants: The theta
parameter of QCD and the cosmological constant of $N=8$ supergravity,''
Nucl.\ Phys.\  B {\bf 176}, 509 (1980).

\bibitem{Hawking1984}
S.W. Hawking,
``The cosmological constant is probably zero,''
Phys. Lett.  B
{\bf 134}, 403 (1984).

\bibitem{Duff1989}
M.J. Duff,
``The cosmological constant is possibly zero, but the proof is probably wrong,''
Phys.\ Lett.\  B {\bf 226}, 36 (1989).

\bibitem{DuncanJensen1989}
M.J. Duncan and L.G. Jensen,
``Four-forms and the vanishing of the cosmological constant,''
Nucl.\ Phys.\  B {\bf 336}, 100 (1990).

\bibitem{BoussoPolchinski2000}
R. Bousso and J. Polchinski,
``Quantization of four-form fluxes and dynamical
neutralization of the  cosmological constant,''
JHEP {\bf 0006}, 006 (2000), arXiv:hep-th/0004134.

\bibitem{Aurilia-etal2004}
A. Aurilia and E. Spallucci,
``Quantum fluctuations of a `constant' gauge field,''
Phys.\ Rev.\  D {\bf 69}, 105004 (2004), arXiv:hep-th/0402096.

\bibitem{Wu2008}
Z.C. Wu,
``The cosmological constant is probably zero, and a proof is
possibly right,''
Phys. Lett.  B {\bf 659}, 891 (2008), arXiv:0709.3314.

\bibitem{Rubakov2001}
V.A.~Rubakov,
``Large and infinite extra dimensions: An introduction,''
Phys.\ Usp.\  {\bf 44}, 871 (2001),
arXiv:hep-ph/0104152.


\bibitem{Lebedev1989}
V.V. Lebedev and A.R. Muratov,
``Dynamics of micelles and vesicles,''
Sov. Phys. JETP {\bf 68}, 1011 (1989).

\bibitem{Kats1993}
E.I. Kats and V.V. Lebedev,
{\it Fluctuational Effects in the Dynamics of Liquid Crystals} (Springer, Berlin, 1993).

\bibitem{Bij1982}
 J.J. van der Bij, H. van Dam, and Y.J. Ng,
 ``The exchange of massless spin two particles,''
Physica {\bf 116} A, 307 (1982).

\bibitem{Zee1983}
A. Zee,
``Remarks on the cosmological constant problem,''
in: S.L. Mintz and A. Perlmutter
(eds.) High-Energy Physics: Proceedings of Orbis Scientiae 1983 (Plenum Press, N.Y., 1985), p. 211.

\bibitem{Dragon1988}
W. Buchm\"uller and N. Dragon,
``Einstein gravity from restricted coordinate invariance,''
Phys. Lett. B {\bf 207}, 292 (1988).


\newpage
\bibitem{Henneaux1989}
M. Henneaux and C. Teitelboim,
``The cosmological constant and general covariance,''
Phys. Lett. B {\bf 222}, 195 (1989).


\bibitem{Klinkhamer2016}
F.R.~Klinkhamer,
``A generalization of unimodular gravity with vacuum-matter energy exchange,''
to appear in Int. J. Mod. Phys. D,  
arXiv:1604.03065. 

\bibitem{Sorkin2013}
M. Ahmed and R.D. Sorkin, ``Everpresent $\Lambda$. II. Structural stability,''
Phys.\ Rev.\  D {\bf 87},  063515 (2013),
arXiv:1210.2589. 

\bibitem{Fursaev1999}
D.V. Fursaev,
``Energy, Hamiltonian, Noether charge, and black holes,''
Phys. Rev. D {\bf 59}, 064020 (1999),
arXiv:hep-th/9809049.

\bibitem{SamuelSinha2006} 
J.~Samuel and S.~Sinha,
``Surface tension and the cosmological constant,''
Phys. Rev. Lett.  {\bf 97}, 161302 (2006);
arXiv:cond-mat/0603804.

\end{thebibliography}
\end{document}